# Observation of spin polarons in a frustrated moiré Hubbard system

Zui Tao[1*], Wenjin Zhao[2*], Bowen Shen[1*], Patrick Knüppel[3], Kenji Watanabe[4], Takashi Taniguchi[4], Jie Shan[1,2,3**], Kin Fai Mak[1,2,3**]

[1]School of Applied and Engineering Physics, Cornell University, Ithaca, NY, USA
[2]Kavli Institute at Cornell for Nanoscale Science, Ithaca, NY, USA
[3]Laboratory of Atomic and Solid-State Physics, Cornell University, Ithaca, NY, USA
[4]National Institute for Materials Science, 1-1 Namiki, 305-0044 Tsukuba, Japan

*These authors contributed equally
**Email: jie.shan@cornell.edu; kinfai.mak@cornell.edu

**The electron's kinetic energy plays a pivotal role in magnetism. While virtual electron hopping promotes antiferromagnetism in an insulator [1], the real process usually favors ferromagnetism [2-6]. But in kinetically frustrated systems, such as hole-doped triangular lattice Mott insulators, real hopping has been shown to favor antiferromagnetism [7-15]. Kinetic frustration has also been predicted to induce a new quasiparticle - a bound state of the doped hole and a spin flip called a spin polaron - at intermediate magnetic fields, which could form an unusual metallic state [9, 12, 13, 16]. However, the direct experimental observation of spin polarons has remained elusive. Here we report the observation of spin polarons in triangular lattice MoTe$_2$/WSe$_2$ moiré bilayers by the reflective magnetic circular dichroism measurements. We identify a spin polaron phase at lattice filling factor between 0.8 – 1 and magnetic field between 2 - 4 T; it is separated from the fully spin polarized phase by a metamagnetic transition. We determine that the spin polaron is a spin-3/2 particle and its binding energy is commensurate to the kinetic hopping energy. Our results open the door for exploring spin polaron pseudogap metals [12, 13, 16], spin polaron pairing [9, 17] and other new phenomena in triangular lattice moiré materials.**

## Main

The Hubbard Hamiltonian describes interplay between the kinetic lattice hopping ($t$) and the on-site Coulomb repulsion ($U$) of electrons in a lattice [18-20]. In the strong interaction limit ($U/t \gg 1$), the solution is a Mott insulator at half-band filling [21], where the localized electrons interact with weak antiferromagnetic (AF) superexchange energy, $J = \frac{4t^2}{U}$ (Ref. [1]). Nagaoka proved that in the limit of infinite $U$, a single hole doping can produce a ferromagnetic (FM) metal from the kinetic energy of the charge carrier [2]. However, Haerter and Shastry subsequently showed that the Nagaoka theorem is not applicable to kinetically (or electronically) frustrated lattices, in which the charge carriers cannot get the full kinetic energy associated with the geometry [7].

An example is illustrated in Fig. 1a for a hole-doped triangular lattice Mott insulator. If the background is fully spin polarized, hole hopping between a given initial and final site involves paths of both even and odd numbers of hops, which carry opposite signs because the holes have a negative hopping amplitude [7]. This results in destructive interference for hole propagation. But if the background is AF, the different hopping paths have distinguishable spin configurations, interference is suppressed, and the kinetic frustration is released. Kinetic frustration has also been predicted to induce a new quasiparticle at intermediate magnetic fields when the Zeeman energy $E_Z$ lies between $J$ and $t$ ($>> J$) [9, 12, 13, 16]. In this regime, the field is sufficient to fully polarize the background spins. But instead of moving in a FM background, the hole moves as a spin polaron (a bound state of the hole and a spin flip) to suppress interference and lower the kinetic energy (Fig. 1b). A gas of spin polarons is expected to be a pseudogap metal that possesses a single-particle and spin gap [9, 12, 13, 16]. In contrast, the Nagaoka ferromagnetism is applicable to electron-doped Mott insulators [10-14, 22], because the doublons have a positive hopping amplitude. Magnetic asymmetry in electron- and hole-doped triangular lattice Mott insulators has been observed in semiconductor moiré materials based on transition metal dichalcogenides (TMDs) [10, 22] and in ultracold atoms [14], but direct evidence of the spin polarons has remained elusive.

Here we report the observation of spin polarons in TMD moiré materials, which have been shown to realize highly tunable triangular lattice Hubbard models [10, 23-30]. Specifically, we choose AB-stacked $MoTe_2/WSe_2$ moiré bilayers with moiré period of about 5 nm [31, 32]. The relatively small moiré period enhances $t$ and $J$, which helps to stabilize the spin polaron phase at temperatures accessible in our experiment (1.6 K). We use the dual-gated device structure illustrated in Fig. 1c to control the carrier density $v$ (in units of the moiré density) and the perpendicular electric field $E$ across the moiré bilayer. We focus on $E$ = 0.5 V/nm, at which the topmost moiré valence bands in two layers remain well separated to realize the single band Hubbard model [31, 32] (the AB stacking with opposite spin alignment in two layers further suppresses layer hybridization). Similar results are observed at other electric fields in this regime (Extended Data Fig. 1). We vary $v$ to introduce holes in $MoTe_2$. The system is a Mott insulator at $v = 1$ from the transport studies [31, 32]. We refer to the system as hole-doped for $v < 1$ and electron-doped for $v > 1$ although the charge carriers are always holes. We probe the sample magnetization as a function of out-of-plane magnetic field $B$ by performing the reflective magnetic circular dichroism (MCD) measurements at the exciton resonances of $MoTe_2$. Details on the device fabrication and MCD measurements are provided in Methods.

**MCD measurements**
Figures 1d and 1e show the filling dependence of the reflectance contrast (RC) and MCD spectra of the moiré bilayer, respectively. We calibrate the filling factor from electrical transport measurements (Extended Data Fig. 2) The MCD at a small magnetic field of 0.6 T is shown as an example (see Extended Data Fig. 3 for MCD spectrum versus magnetic fields). In the energy range of 1.1 – 1.2 eV, three resonances corresponding to the intralayer moiré excitons in $MoTe_2$ can be identified. These features are enhanced in both the RC and MCD near the Mott insulator state. Similar behavior has been reported in other TMD moiré systems [10, 25, 29]. In order to analyze MCD as a function of tuning parameters, such as the lattice filling factor, magnetic field and temperature, we integrate the MCD modulus over all three exciton resonances (the dashed lines denoting the spectral range for integration). The integrated MCD (referred to simply as MCD below) reflects the difference in occupancy between the K and K' valleys in $MoTe_2$ (Ref. [10, 15, [22]). Because of spin-valley locking, the signal is proportional to the out-of-plane magnetization $M$ (Ref. [29]). But the proportionality coefficient (sensitivity) is filling dependent, and the MCD at different filling factors cannot be directly compared.

Figure 2a illustrates the magnetic-field dependence of MCD for representative filling factors at 1.6 K. While the MCD increases from zero at zero field and saturates at high fields for all filling factors, asymmetry in the behavior for electron and hole doping is apparent. The electron-doped Mott insulator ($\nu$ = 1.1) behaves similarly to the Mott insulator ($\nu$ = 1); the MCD increases with field continuously till reaching full saturation at about 2 T. In contrast, for hole doping ($\nu$ < 1) the MCD shows a plateau at intermediate field between 2 - 4 T before reaching full saturation. This is better illustrated in Fig. 2b with the numerical field derivative of the MCD, which is proportional to the differential magnetic susceptibility, $\chi(B) = \frac{dM}{dB}$. The intermediate MCD plateau manifests a local minimum in $\chi(B)$, which can be identified for $0.8 \lesssim \nu \lesssim 0.95$ (Extended Data Fig. 4). As $\nu$ approaches 1, the plateau MCD becomes increasingly close to the fully saturated MCD and is no longer discernable. At lower filling factors ($\nu$ = 0.74), the intermediate plateau is smeared out, resulting in a continuously increasing MCD till reaching full saturation around 4 T.

The intermediate magnetization plateau for hole doping is observable only at low temperatures. Figure 3a illustrates an example for $\nu$ = 0.91 (see Extended Data Fig. 5 for the corresponding $\chi(B)$). As temperature increases, the plateau becomes increasingly blurred and indiscernible above about 5 K. As a comparison, Fig. 3b shows the same measurement for the Mott insulator. As temperature increases, the magnetic susceptibility decreases and the MCD saturates at increasingly larger fields. The high temperature behavior is general for all filling factors including $\nu$ = 0.91 in Fig. 3a.

**Kinetic magnetism**
The MCD measurements show strong magnetic asymmetry in electron and hole doping of the triangular lattice Mott insulator. We can extract the magnetic interaction for different filling factors by analyzing the temperature dependence of the zero-field MCD slope ($\propto \chi$) [10, 15, 22]. Figure 3c shows two examples with $\nu$ = 0.84 and 0.97. The inverse MCD slope follows the Curie-Weiss law, $\chi^{-1} \propto T - \theta$ (solid lines), for temperatures exceeding $\theta$. Here $\theta$ denotes the Curie-Weiss temperature, which is proportional to the effective spin-spin exchange interaction [33]. A negative (positive) $\theta$ reflects an AF (FM) exchange interaction. The filling factor dependence of $\theta$ is summarized in Fig. 3d. The interaction is AF for the entire filling range of 0.74 - 1.2. While $|\theta|$ is about 2 K for the Mott insulator and remains small for electron doping, it increases linearly with density $x$ for hole doping ($\nu$ = 1 - $x$).

The result is consistent with kinetic magnetism in frustrated lattices discussed above and observed in other TMD moiré systems [7-15, 22]. The observed AF interaction with a small $|\theta|$ for the Mott insulator reflects a strongly interacting system. The AF interaction releases the kinetic frustration for doped holes. The effective exchange interaction hence scales with kinetic hopping and hole density [13] ($|\theta| \propto x|t|$), as observed in Fig. 3d. Such kinetic antiferromagnetism from hole doping is suppressed at commensurate fractional fillings, where generalized Wigner crystals emerge [24, 34, 35] and the doped holes are localized by long-range Coulomb interactions (Ref. [15] and Extended Data Fig. 6). Note that the magnetic interaction does not turn FM immediately above $\nu = 1$ as is expected for the infinite $U$ limit [2, 13]. This is likely caused by the larger bandwidth of the upper Hubbard band [34, 36], which reduces the effective interaction strength and delays the onset of the FM interaction [37]. Disorder and long-range Coulomb interaction could also hamper the onset of the FM interaction.

**Spin polarons**

The main finding of our experiment is the presence of a magnetization plateau at field between 2 - 4 T at low temperatures and for hole doping down to $x \approx 0.2$. The observation is fully compatible with the predicted spin polaron phase [9, 12, 13, 16]. In this picture, as magnetic field increases, the spins in the AF background surrounding the doped holes are first aligned and the magnetization increases. Above the saturation field $B_s$ at which the Zeeman energy overcomes the AF spin-spin interaction, spin polarons rather than bare holes in a FM background form to release the kinetic frustration. Naturally, the binding energy of a spin polaron is on the order of kinetic hopping $t$ (~ $0.5t$ from calculations for the triangular lattice Mott insulator with one doped hole in the large $U/t$ limit [12, 13, 16]). A further increase of the field eventually unbinds the spin polarons and induces full magnetization. The metamagnetic transition occurs at field $B_m$, at which the Zeeman energy is comparable to $t$. Alternatively, the spin polarons can be unbound by thermal excitations and are hence observable only at temperatures much lower than $t$. In addition, because kinetic frustration applies only to hole doping in the triangular lattice Mott insulator [7-15], we observe the intermediate magnetization plateau for hole doping but not electron doping. As hole density increases, the spin polaron wavefunctions start to overlap, the magnetization plateau shrinks and is eventually smeared out. The observed upper density limit ($x \approx 0.2$) is in good agreement with the numerical result of $x \approx 0.15$ for the large $U/t$ limit [9, 12, 13, 16].

In the fully spin polarized phase, the magnetization is proportional to $\frac{1}{2}(1 - x)$ since each charge carrier has a spin of ½. In the spin polaron phase, the magnetization is further reduced by $x$ to yield $\left(\frac{1}{2} - \frac{3}{2}x\right)$ since each spin flip changes the spin by -1. The spin polaron therefore carries a spin of 3/2. This is verified in our experiment by examining the ratio of the intermediate magnetization plateau to the saturation magnetization (Fig. 4a and Extended Data Fig. 7). Here the intermediate magnetization plateau is measured at the field corresponding to the local minimum of $\chi(B)$ in Fig. 2b. The ratio follows the expected doping dependence of $\frac{1-3x}{1-x}$ (dashed line) for a spin-3/2 quasiparticle [12, 13, 16].

We summarize in Fig. 4b the spin polaron phase in lattice filling factor and magnetic field extracted from our experiment at 1.6 K. The spin polarons are observed at filling factor between $0.8 - 1$ and magnetic filed between $B_s$ and $B_m$. The two fields or edges of the intermediate magnetic plateau are marked by the dashed lines in Fig. 2. We take the saturation field $B_s$ to be the one for the largest curvature of MCD (or the steepest drop of $\chi$). We take the metamagnetic transition field $B_m$ to be the one for the steepest rise of MCD (or the local maximum of $\chi$). Above $B_m$, the system is fully spin polarized.

The relevant energy scales can be estimated from the observed $B_s \sim 2$ T and $B_m \sim 4$ T. In the low-temperature limit, the saturation field for the undoped Mott insulator is set by the AF superexchange interaction [12], $g\mu_B B_s = \frac{9}{2}J$, where $\mu_B$ denotes the Bohr magneton and $g \approx 10$ is the spin $g$-factor for the topmost valence band of TMDs [38]. This yields $J \sim 0.25$ meV, which is consistent with the extracted Curie-Weiss temperature of $\sim 2$ K. On the other hand, the metamagnetic transition is set by the spin polaron binding energy [12, 13, 16], $g\mu_B B_m \approx 0.5t$. This yields $t \sim 5$ meV, which far exceeds $J$ as expected in the large $U/t$ limit. The estimated bandwidth ($9t \sim 45$ meV) is also in good agreement with ab initio band structure calculations for AB-stacked $MoTe_2/WSe_2$ (Ref. [39]). The measured spin polaron binding energy ($\sim 2.5$ meV) provides a measure of the spin gap of the predicted spin polaron pseudogap metal [16].

We note that the observed magnetization plateau is generally broadened compared to the theoretical predictions [13, 16]. We attribute the discrepancy to multiple factors, including the thermal excitations at 1.6 K, the finite $U/t$, and potentially also the long-range Coulomb repulsion. The long-range Coulomb repulsion is known to be important in our moiré system, as manifested in the presence of Wigner-Mott insulators at fractional fillings [24, 34, 35], but its effect on the stability of the spin polaron phase remains open. Furthermore, for relatively large hole densities, theory has also predicted spin polaron bound states that involve multiple spin flips (for instance, bipolarons) which give rise to multiple magnetization plateaus before reaching full saturation [9]. This is expected to broaden the intermediate magnetization plateau.

**Conclusions**
In conclusion, we have provided direct experimental evidence for the spin polaron phase in a hole-doped triangular lattice Mott insulator subjected to an intermediate magnetic field between 2 – 4 T. The observed intermediate magnetization plateau, its ratio to the saturation magnetization as well as doping and temperature dependences are all in good agreement with the predicted spin polaron phase in the large $U/t$ limit [9, 12, 13, 16]. The spin polaron phase also correlates with the observed enhanced AF interaction at zero field since both are of the origin of kinetic frustration. Our results pave the way to establish a spin polaron metal [12, 13, 16] and explore the possibility of spin polaron mediated superconductivity [9, 17] in TMD moiré materials.

**Methods**
**Device fabrication**
Hall bar devices of the AB-stacked $MoTe_2/WSe_2$ moiré bilayers were fabricated using the reported layer-by-layer dry transfer method [31]. In short, thin flakes of hexagonal boron nitride (h-BN), graphite, $MoTe_2$ and $WSe_2$ were first exfoliated from bulk crystals onto silicon substrates, and flakes of appropriate thickness were identified according to their reflectance contrast under an optical microscope. We also determined the crystal orientations of $MoTe_2$ and $WSe_2$ monolayers by angle-resolved optical second-harmonic generation before stacking [10, 24]. A polycarbonate (PC) stamp was used for dry transfer. The TMD moiré bilayers are encapsulated in graphite/h-BN top and bottom gates. A relatively thin h-BN layer (~ 5 nm) was used in the top gate to achieve out-of-plane electric field up to ~1 V/nm. Platinum electrodes of ~ 5 nm thickness were pre-deposited onto the bottom gate h-BN to achieve good electrical contact to the moiré bilayers while keeping the strain effects minimal.

**Reflective MCD measurements**
The optical measurements were performed in a closed-cycle helium cryostat equipped with a superconducting magnet (Attocube, Attodry 2100, base temperature 1.6 K). A superluminescent diode (Exalos, EXS210007-01) of peak wavelength 1070 nm and full-width-at-half-maximum (FWHM) bandwidth 90 nm was used as the light source. The output of the diode was coupled to a single mode fiber and focused onto the device under normal incidence by a low temperature microscope objective (0.8 numerical aperture). A combination of a linear polarizer and an achromatic quarter-wave plate was used to generate the left ($\sigma^+$) and right ($\sigma^-$) circularly polarized light. The reflected light of a given helicity was spectrally resolved by a spectrometer coupled to a liquid nitrogen cooled InGaAs one-dimensional array sensor. The reflectance contrast spectrum was obtained by comparing the reflected light spectrum from the sample to the reference spectrum measured on a heavily doped device (which is featureless in the spectral region of interest). The incident intensity on the sample was below 50 nW/$\mu$m$^2$, which was chosen to practically eliminate the effect of illumination on the magnetic properties of the samples.

The MCD spectrum is defined as $\frac{R^+ - R^-}{R^+ + R^-}$, where $R^+$ and $R^-$ denote the reflection intensity of the left and right circularly polarized light. Three intralayer moiré exciton resonances from MoTe$_2$ can be identified. We integrate the MCD modulus over a fixed spectral range of 1.12 - 1.17 eV, which covers all three intralayer moiré exciton resonances.

**Acknowledgement**
We thank Liang Fu, Yang Zhang and Margarita Davydova for many fruitful discussions.

# Figures

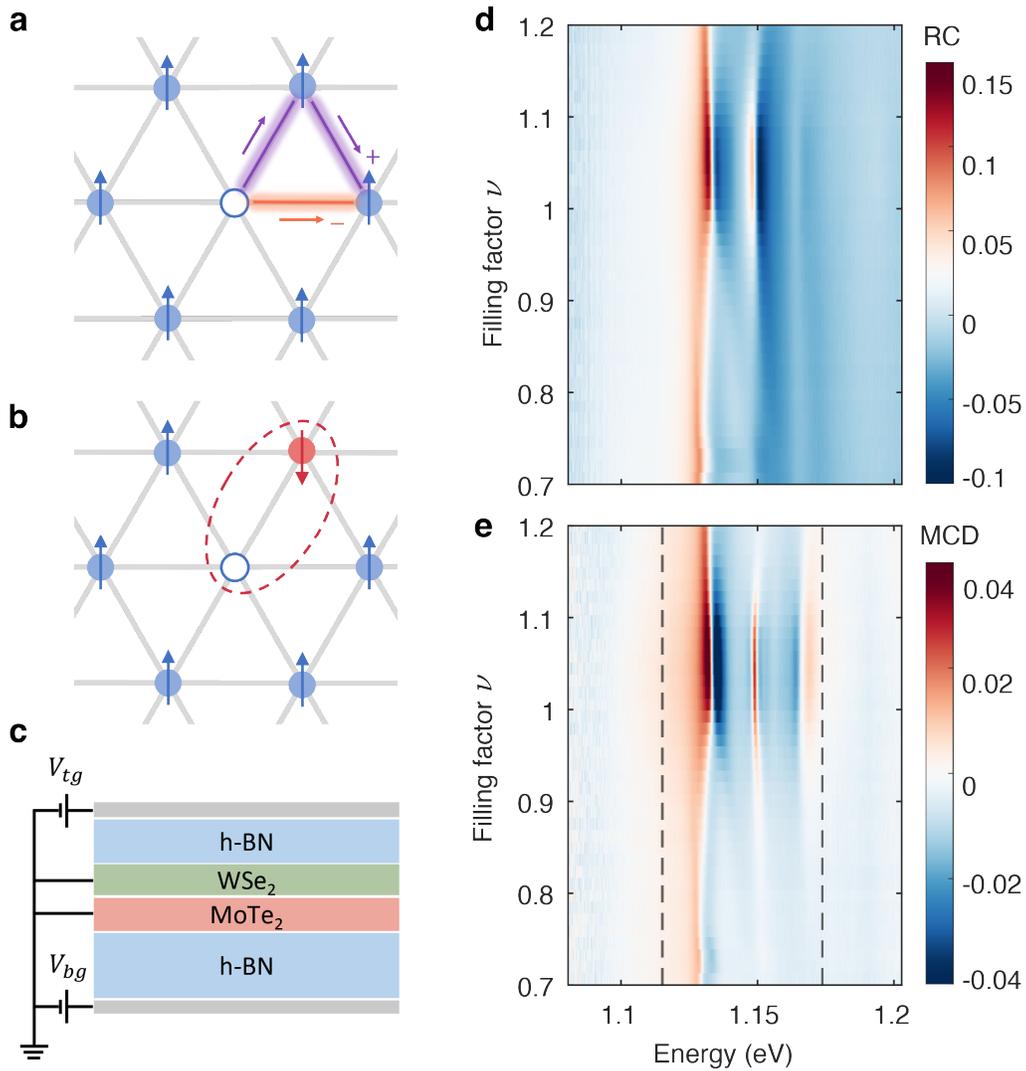

**Figure 1 | Experimental setup. a,** Kinetic frustration of a hole (white dot) in a fully spin polarized background (blue dots with arrows denoting the spin alignment) on a triangular lattice. The purple path involves two hops and carries a positive sign, and the orange path involves one hop and carries a negative sign; they interfere destructively. **b,** The propagation of a spin polaron (bound state of a doped hole and a spin flip, denoted by the dashed ellipse) is not frustrated. **c,** Schematic side view of a dual-gated AB-stacked MoTe$_2$/WSe$_2$ moiré bilayer with graphite/h-BN gates. The sample is grounded. The top and bottom gate voltage ($V_{tg}$ and $V_{bg}$) are applied to control the lattice filling factor and out-of-plane electric field. **d,e,** Optical reflectance contrast (RC) (**d**) and MCD spectrum (**e**) at B = 0.6 T as a function of filling factor $\nu$. The out-of-plane electric field is fixed at 0.5 V/nm. Three intralayer moiré exciton resonances from MoTe$_2$ can be identified. The dashed lines show the spectral range for MCD integration.

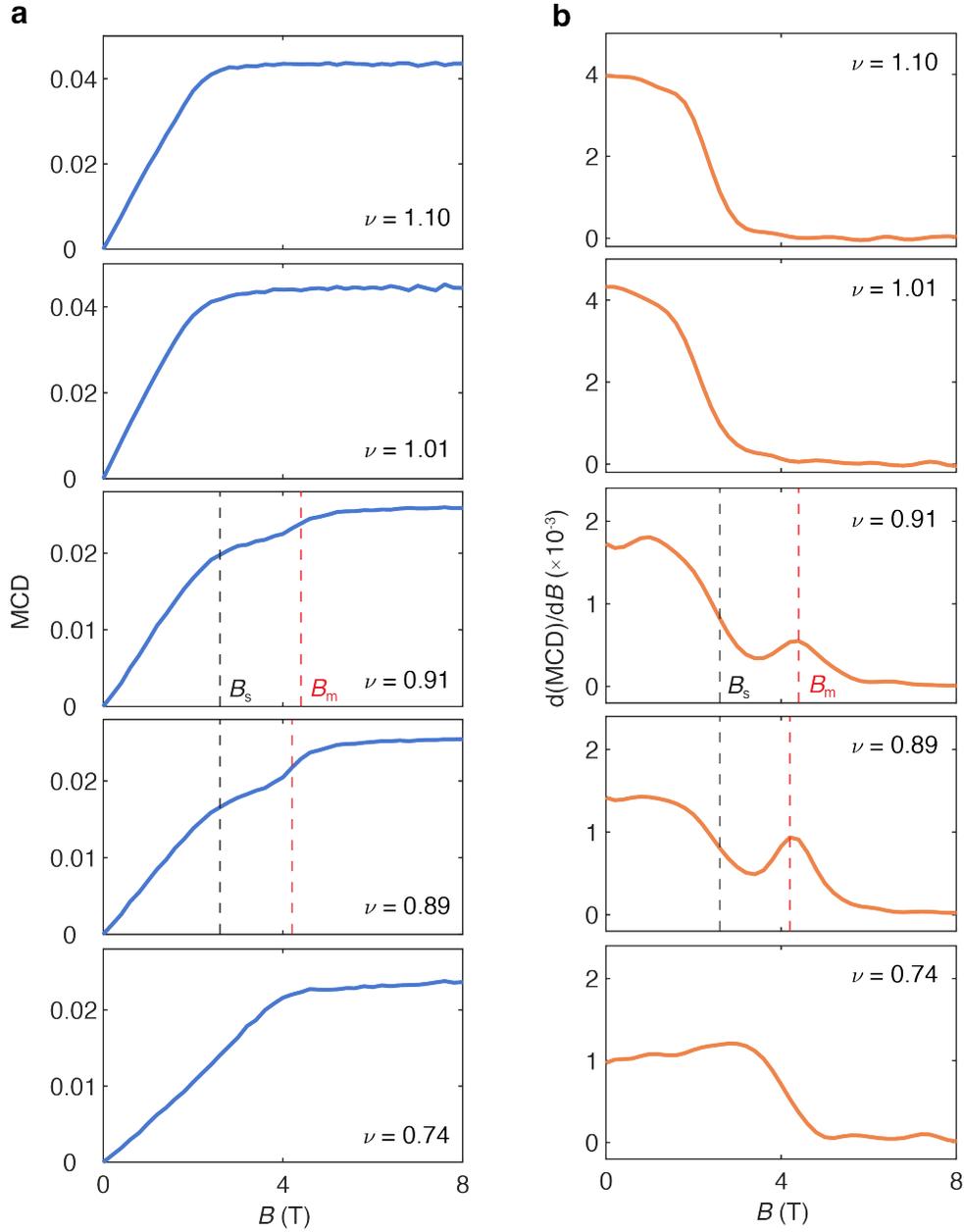

**Figure 2 | Intermediate magnetization plateau. a,b,** Magnetic-field dependence of the MCD (proportional to the magnetization, **a**) and MCD derivative (proportional to the differential susceptibility, **b**) at representative filling factors. An intermediate MCD plateau at magnetic field between 2 – 4 T is observed for filling factor between 0.8 – 1. Two examples are shown for $\nu$ = 0.91 and 0.89. The intermediate MCD plateau manifests a local minimum in the MCD derivative. The dashed lines denote the two ends of the plateau, the saturation field $B_s$ and the metamagnetic transition field $B_m$. The saturation field is determined by the steepest drop in the MCD derivative, and the metamagnetic transition field is determined by the local maximum of the MCD derivative. For filling above 1 and below 0.8, only one of the fields can be identified.

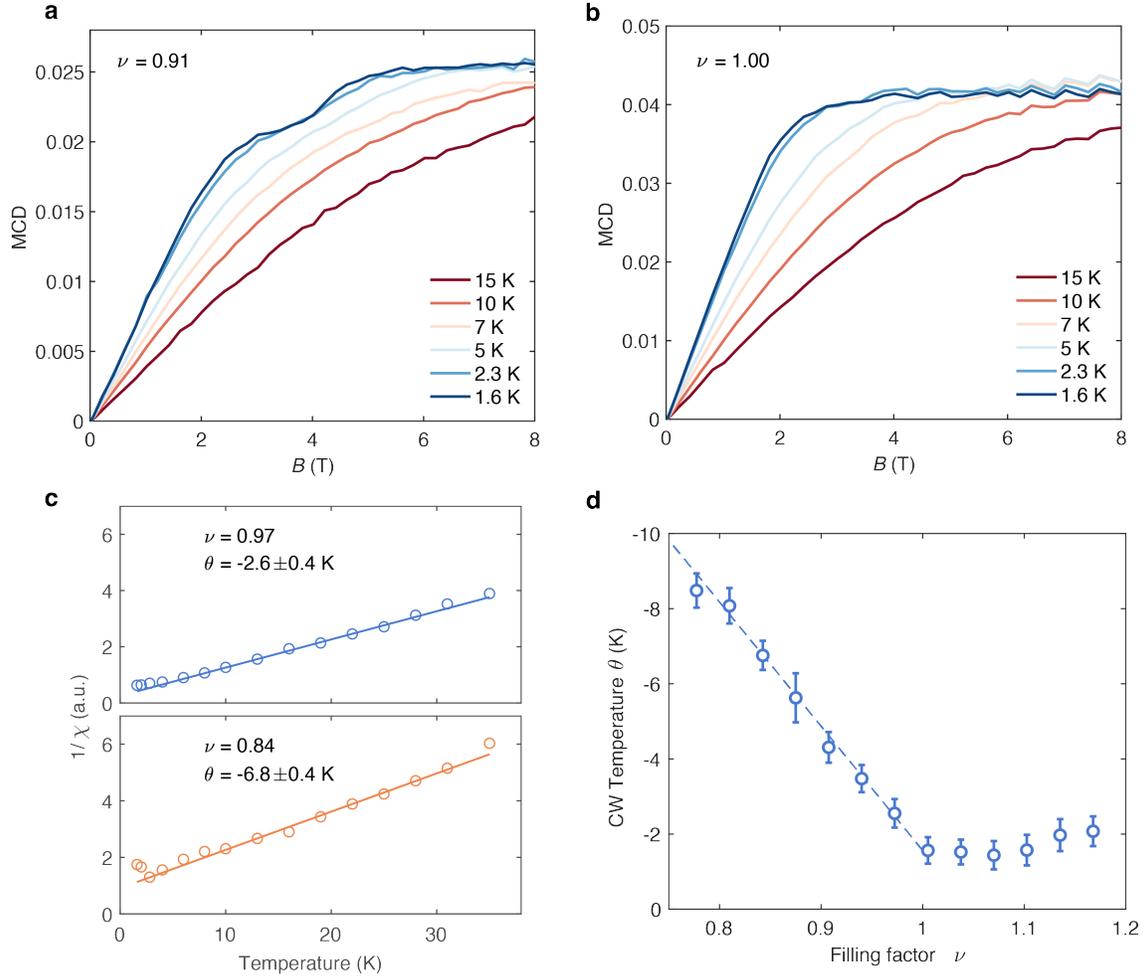

**Figure 3 | Kinetic magnetism and kinetic frustration. a,b,** Magnetic-field dependence of the MCD for $\nu = 0.91$ (**a**) and $\nu = 1.00$ (**b**) at varying temperatures. **c,** Temperature dependence of the zero field slope of the MCD (symbols) and the Curie-Weiss analysis (lines) at two representative filling factors, where $\theta$ is the extracted Curie-Weiss temperature. **d,** Filling dependent $\theta$ (symbols). The error bars represent the uncertainty of the Curie-Weiss analysis. For the entire filling range, the interaction is AF ($\theta < 0$). Specifically, $|\theta|$ is small for the Mott insulator and there is asymmetry between electron and hole doping. For electron doping, $|\theta|$ remains small. For hole doping, $|\theta|$ increases linearly with hole density with slope of about 33 K (dashed line).

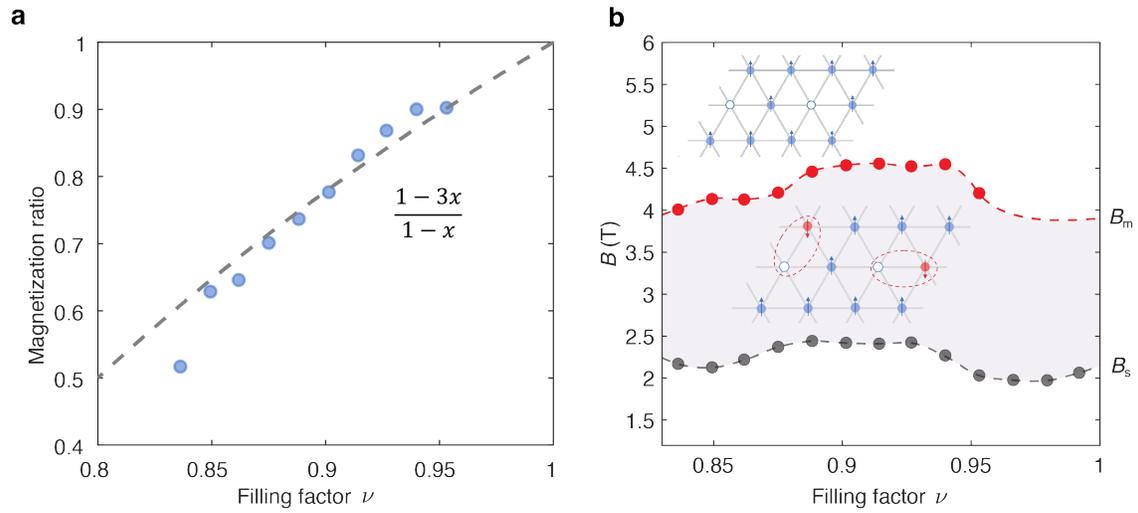

**Figure 4 | Spin polaron phase. a,** Filling dependent ratio of the intermediate magnetization plateau to the saturation magnetization (symbols). It follows $\frac{1-3x}{1-x}$ (dashed line), where $x$ is the hole doping density. **b**, The spin polaron phase (filled area) is observed at filling factor between $0.8 - 1$ and magnetic field between the saturation field $B_s$ and the metamagnetic transition field $B_m$ (symbols). Above $B_m$, the system is fully spin polarized. The dashed lines are guide to the eye. The insets are schematics of the two phases, where the white, blue and red dots represent doped holes, charge carriers with aligned spin and charge carriers with flipped spin. The spin alignment is shown by the arrows.

**Extended Data Figures**

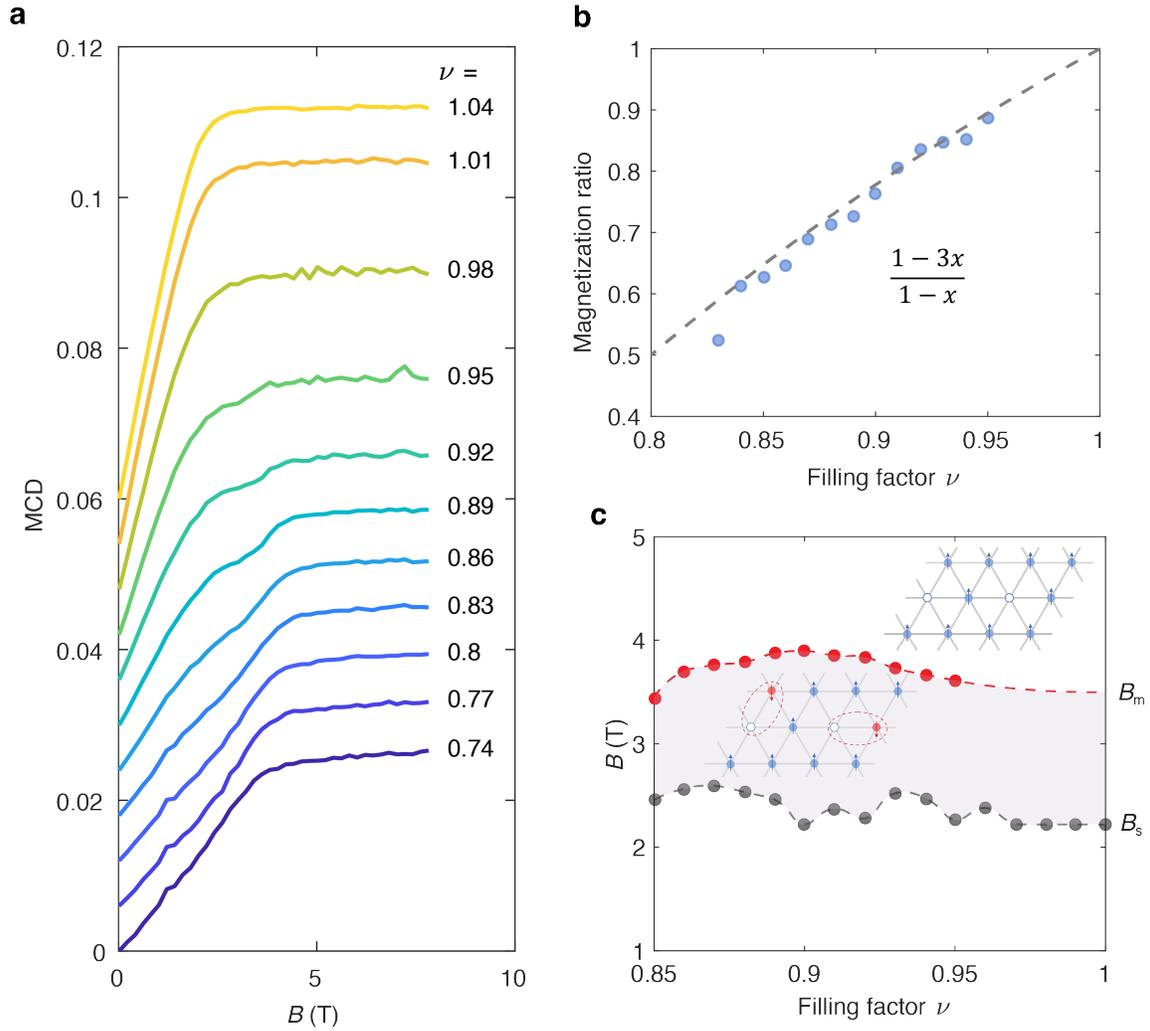

**Extended Data Figure 1 | Additional results at $E$ = 0.3 V/nm. a,** Magnetic-field dependent MCD at representative filling factors at 1.6 K. The curves are vertically displaced for clarity. The intermediate MCD plateau is observed for $0.83 \lesssim \nu \lesssim 0.95$. **b,** Filling factor dependent magnetization ratio between the intermediate plateau and full saturation. The dashed line describes $\frac{1-3x}{1-x}$. **c,** Filling factor dependent saturation field $B_s$ and metamagnetic transition field $B_m$ extracted from experiment (symbols). The lines are guide to the eye. The shaded area represents the region for the spin polaron phase.

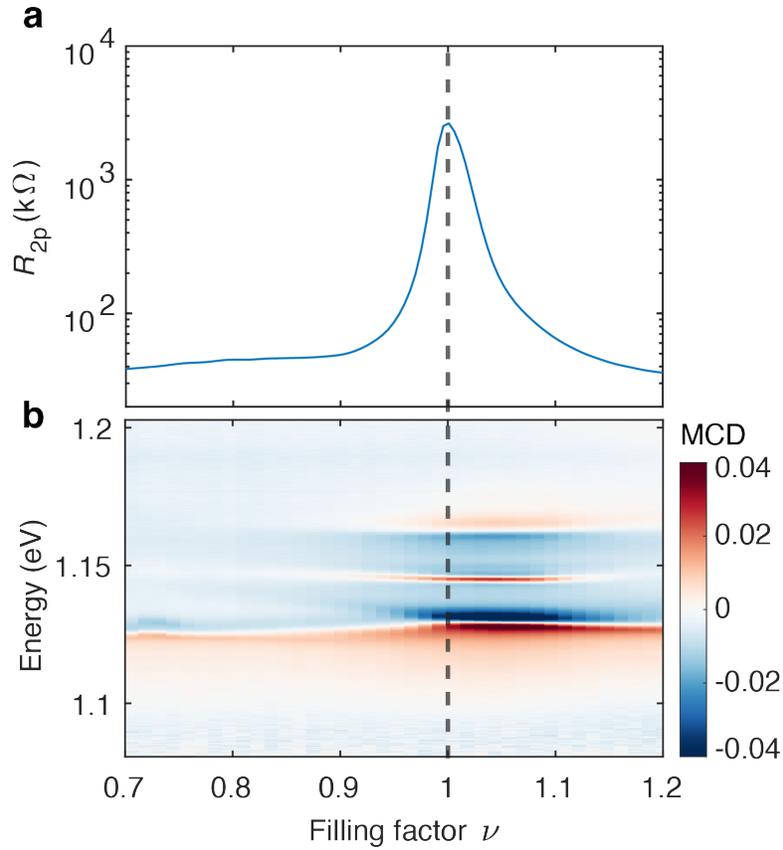

**Extended Data Figure 2 | Filling factor calibration. a,b,** Filling factor dependence of the two-terminal longitudinal resistance (**a**) and the MCD spectrum (at 0.6 T, **b**) at 1.6 K. The resistance peak identifies the $\nu = 1$ Mott insulating state (dashed line).

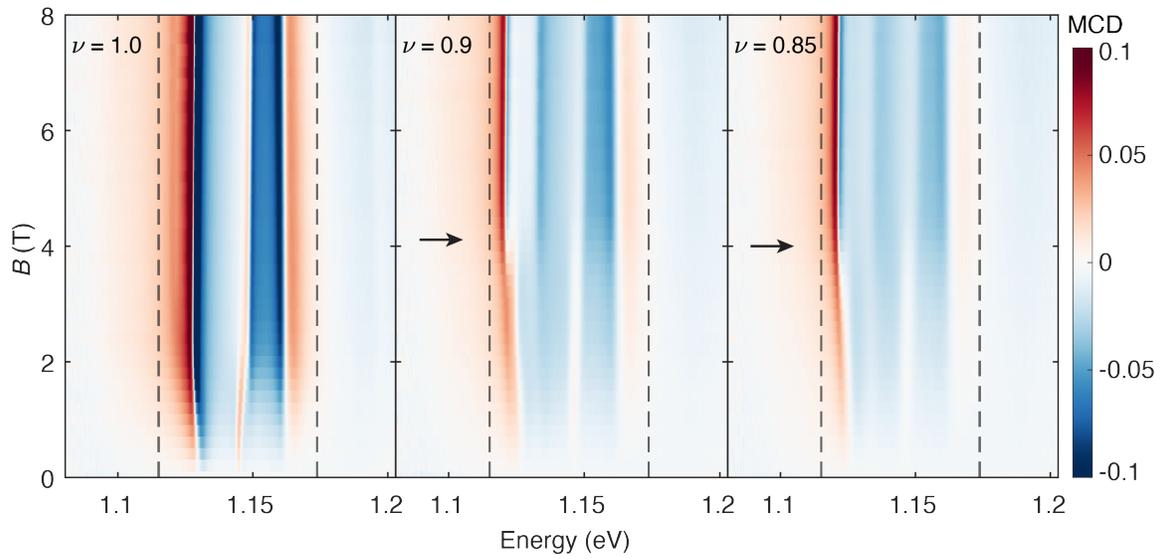

**Extended Data Figure 3 | Magnetic field dependent MCD spectrum at selected filling factors.** The electric field is fixed at 0.5 V/nm; the temperature is 1.6 K. Three intralayer moiré exciton resonances from $MoTe_2$ can be identified. The dashed lines show the spectral range for MCD integration. The arrows indicate the metamagnetic transition field $B_m$, where a kink in the MCD spectrum is observed.

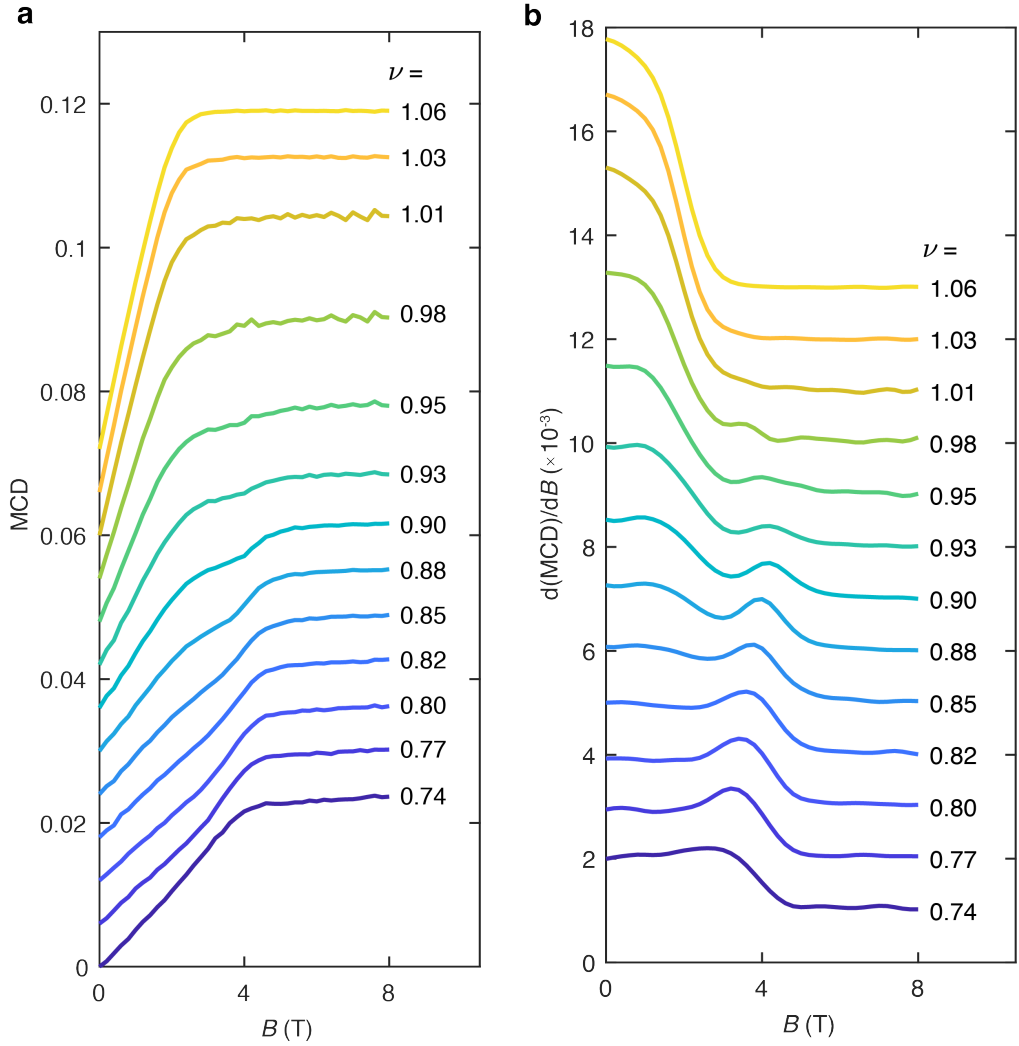

**Extended Data Figure 4 | Magnetic field dependent MCD at varying fillings. a,b,** Magnetic field dependence of the MCD (proportional to the magnetization, **a**) and the MCD derivative (proportional to the differential susceptibility, **b**) at representative filling factors and at 1.6 K. The electric field is fixed at 0.5 V/nm. The curves are vertically displaced for clarity. An intermediate MCD plateau at magnetic field between 2 - 4 T is observed for filling factor between 0.8 – 1. The MCD plateau manifests as a local minimum in MCD derivative.

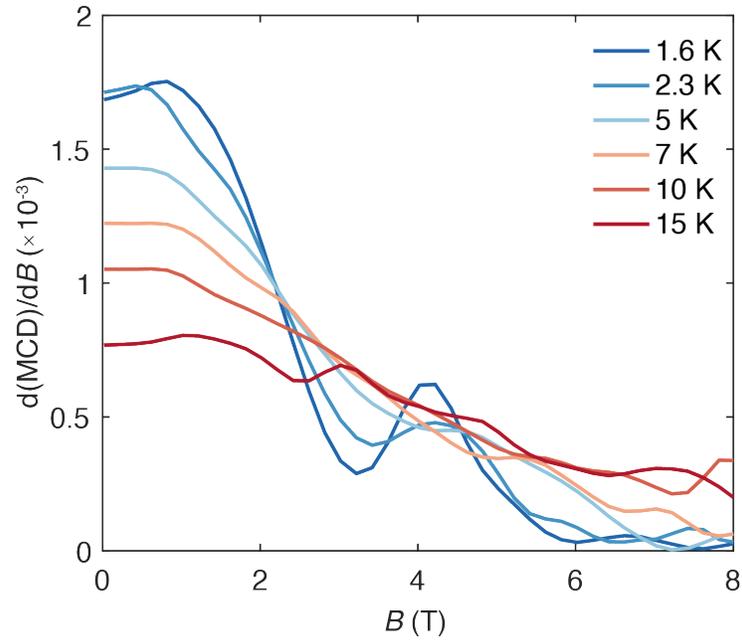

**Extended Data Figure 5 | Magnetic-field dependence of the MCD derivative for $\nu = 0.91$ at varying temperatures.** The numerical derivatives are taken from the data in Fig. 3a. The intermediate MCD plateau in Fig. 3a manifests a local minimum in the MCD derivative, which can only be observed below about 5 K.

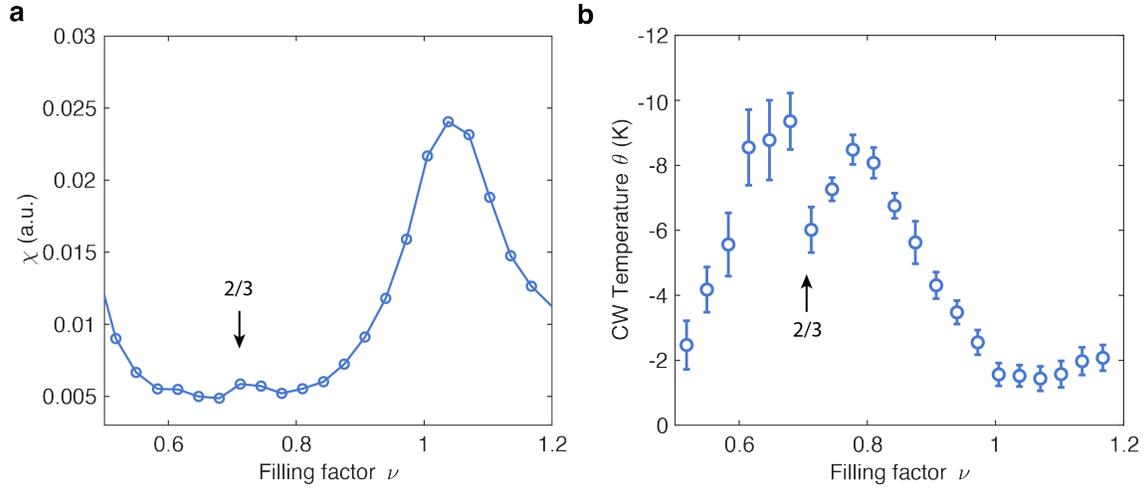

**Extended Data Figure 6 | Filling factor dependent susceptibility and Curie-Weiss temperature $\theta$. a,b,** Filling factor dependence of the magnetic susceptibility $\chi$ at 1.6 K (**a**) and the Curie-Weiss temperature $\theta$ (**b**). The magnetic susceptibility is obtained by fitting the MCD slope at zero magnetic field. The analysis is described in the main text, and the result near $\nu = 1$ is included in Fig. 3d. A suppression of the AF interaction is observed, as evidenced by a peak in $\chi$ and a dip in $\theta$ for the Wigner-Mott insulator at $\nu = 2/3$ (arrows).

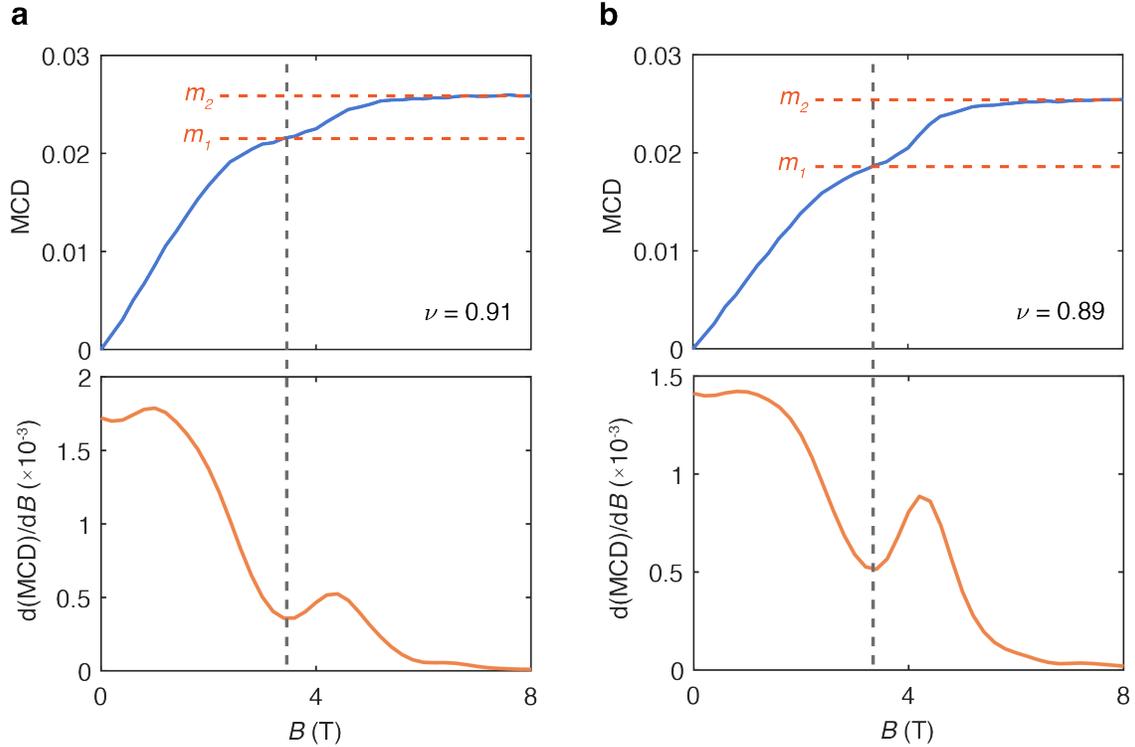

**Extended Data Figure 7 | Analysis of the magnetization ratio. a,b,** Examples of determining the magnetization ratio at $\nu = 0.91$ (**a**) and $\nu = 0.89$ (**b**). The magnetization ratio is obtained by the MCD ratio between the spin polaron magnetization $m_1$ and the saturation magnetization $m_2$, i.e. $\frac{m_1}{m_2}$ (horizontal dashed lines). The spin polaron magnetization $m_1$, which corresponds to the intermediate plateau in MCD, is determined by the local minimum of the MCD derivative (vertical dashed lines).